\documentclass[conference]{IEEEtran}

\usepackage{cite}

\ifCLASSINFOpdf
  \usepackage[pdftex]{graphicx}
  \graphicspath{{figs/}}
  \DeclareGraphicsExtensions{.pdf,.jpeg,.png}
\else
  \usepackage[dvips]{graphicx}
  \graphicspath{{eps/}}
  \DeclareGraphicsExtensions{.eps}
\fi

\usepackage{amsmath}
\usepackage{array}

\ifCLASSOPTIONcompsoc
  \usepackage[caption=false,font=normalsize,labelfont=sf,textfont=sf]{subfig}
\else
  \usepackage[caption=false,font=footnotesize]{subfig}
\fi

\usepackage{dblfloatfix}

\usepackage{url}

\usepackage{amsthm}
\usepackage[titlenumbered,ruled]{algorithm2e}
\usepackage{multirow}
\DeclareMathOperator*{\argmin}{arg\,min}
\DeclareMathOperator*{\argmax}{arg\,max}

\begin{document}

\title{Integrate Multi-omic Data Using Affinity Network Fusion (ANF) for Cancer Patient Clustering}

\author{\IEEEauthorblockN{Tianle Ma}
\IEEEauthorblockA{Department of Computer Science and Engineering\\
University at Buffalo (SUNY)\\
Buffalo, New York 14260-2500\\
Email: tianlema@buffalo.edu}
\and
\IEEEauthorblockN{Aidong Zhang}
\IEEEauthorblockA{Department of Computer Science and Engineering\\
	University at Buffalo (SUNY)\\
	Buffalo, New York 14260-2500\\
	Email: azhang@buffalo.edu}}

\maketitle

\begin{abstract}
Clustering cancer patients into subgroups and identifying cancer subtypes is an important task in cancer genomics. Clustering based on comprehensive multi-omic molecular profiling can often achieve better results than those using a single data type, since each omic data type (representing one view of patients) may contain complementary information. However, it is challenging to integrate heterogeneous omic data types directly. Based on one popular method -- Similarity Network Fusion (SNF), we presented Affinity Network Fusion (ANF) in this paper, an ``upgrade'' of SNF with several advantages. Similar to SNF, ANF treats each omic data type as one view of patients and learns a fused affinity (transition) matrix for clustering. We applied ANF to a carefully processed harmonized cancer dataset downloaded from GDC data portals consisting of 2193 patients, and generated promising results on clustering patients into correct disease types. Our experimental results also demonstrated the power of feature selection and transformation combined with using ANF in patient clustering. Moreover, eigengap analysis suggests that the learned affinity matrices of four cancer types using our proposed framework may have successfully captured patient group structure and can be used for discovering unknown cancer subtypes.  
\end{abstract}

\ifCLASSOPTIONpeerreview
\begin{center} \bfseries EDICS Category: 3-BBND \end{center}
\fi

\IEEEpeerreviewmaketitle

\section{Introduction}
Cancer genomics projects such as TCGA have generated comprehensive multi-omic molecular profiling for dozens of cancer types. Mining huge amounts of omic data to discover cancer subtypes and disease mechanisms is still a hot topic. 

As patients with cancer from the same primary sites, for example, lung cancer, can be very different from each other in terms of disease progression, response to treatments, etc. One important task is to further cluster cancer patients of the same cancer type into subgroups and define new cancer subtypes with comprehensive molecular signatures associated with distinct clinical features. 

There are many challenges to cluster patients into subgroups since cancer patients are very heterogeneous. Even though we can always cluster patients into groups by running a specific clustering algorithm, the clustering results may not be robust, i.e., slightly changing clustering methods or parameter settings may lead to a different clustering result. Moreover, there is no groundtruth to decide which methods and parameter settings work best. While the omic data collected are comprehensive, they are heterogeneous and noisy, too. If we use each type of omic data to cluster patients, we can probably generate different results. Is it possible to generate a robust clustering result and use ``groundtruth'' to justify it? This paper mainly focus on this problem.  

Since each type of omic data may contain some complementary information about the patients and disease, we can perform integrative network analysis with multi-omic data. Many methods that have been developed to integrate multi-omic data for patient clustering in the past several years are either based on probabilistic models or network models \cite{Bersanelli2016}. It has been demonstrated that patient clustering based on similarity network fusion (SNF) \cite{Wang2014} can usually achieve promising results compared with other methods such as iCluster \cite{Shen2012} or KMeans. While SNF works well in clustering patients, we find that the required computational operations in SNF can be significantly reduced and simplified to get a reliable fused affinity network. Based on SNF, in this paper we presented Affinity Network Fusion (ANF) with several advantages compared with SNF.

Our contribution can be summarized as follows:

First, ANF presented here can be seen as an improved version of SNF with several advantages. ANF requires much less computation while generating as good as or even better results than those from SNF. ANF can incorporate weights of each view, while SNF is unweighted. Moreover, ANF has a much more clearer interpretation, while SNF contains some ``mysterious'' operations.
 
Second, we cleverly selected a ``gold'' dataset with true class labels for cancer patients clustering problem. Thus instead of using internal clustering evaluation metrics, we were able to use external metrics to evaluate clustering methods. 
We used the newest release of harmonized cancer datasets from Genomic Data Commons Data Portal (since the data is harmonized, it is of high quality for large-scale integration), and carefully selected 2193 cancer patients from four primary sites with known disease types. With this dataset, we were able to demonstrate the power of affinity network fusion technique in cancer patient clustering. Our experimental results also showed survival information alone may not be sufficient for evaluating disease (sub)type discovery methods. 

Third, using harmonized gene expression and miRNA expression data, we were able to demonstrate that feature selection and transformation of ``raw'' counts or other genomic features could lead to better clustering results. Specifically, we find that log transformation or variance stabilizing transformation \cite{Durbin2002} of raw counts data usually perform better than directly using normalized expression values such as FPKM values of gene expression.

Forth, the experimental results on this relatively large dataset (2193 cancer patients with gene expression, miRNA expression and DNA methylation data) are very promising. The learned fused affinity matrices for the selected four cancer types matched well with both true class labels and the theory of spectral clustering based on eigengap analysis (Sec.~\ref{sec:eigengap}), which can be reliably used for unknown cancer subtype discovery and identifying subtype-specific molecular signatures. 

The rest of paper is organized as follows. In section \ref{sec:related_work}, we briefly summarize some related work. In section \ref{sec:general_framework}, we briefly describe a general framework for clustering complex objects. In section \ref{sec:anf}, we describe ANF framework in detail. In section \ref{sec:res}, we presented experimental results and analysis. The last section gives conclusion.

\section{Related Work} \label{sec:related_work}

Integrating multi-omic data has been a hot topic in recent years. There are multiple good reviews on this topic \cite{Bersanelli2016,Meng2016}. Various techniques can be classified into four groups based on whether they are probabilistic or network based: Non-Probabilistic Non-Network, Non-Network Probabilistic, Non-Probabilistic Network, and Probabilistic Network. 

Non-Probabilistic Non-Network approaches do not assume a probability distribution or use graph theory to integrate multi-omic data. Instead direct regression or correlation analysis are applied to multi-omic data. For example, partial least square can identify the features that are most informative for predicting clinical outcomes, and can weigh different sources. Canonical Correlation Analysis leverages correlations among different features in various sources to select the most informative features. Some frameworks apply Expectation-Maximization approaches to learn the weight of each source in the optimization framework.

Non-Network Probabilistic approaches refer to methods that employs a probabilistic approach containing latent variables with a prior distribution. For instance, iCluster \cite{Shen2012}, a widely used cancer subtype clustering method, assumes different types of data share a common latent feature space that can be learned through Expectation-Maximization (EM). Clustering is performed on the learned latent feature space.

Non-probabilistic network approaches often do not assume a prior distribution of a set of latent variables. In fact, many models of this category even do not have latent variables. Most of them try to leverage interaction network to diffuse the signal and fuse various networks. Typical examples include HotNet2 \cite{Leiserson2015} and SNF \cite{Wang2014}. HotNet2 diffuses genetic mutation signals through gene interaction network to get a smoothed mutation profile, which is then used for detecting ``hot spot'' of gene subnetworks that might be disease-causing. SNF constructs patient similarity networks using different types of data. The novelty of SNF is that it fuses different patient similarity networks to achieve a consensus similarity network that could be more reliable instead of directly combing heterogeneous features. The fused patient similarity network is then used for clustering patients into disease subtypes.

Probabilistic network approaches usually employs Bayesian approaches, which usually incorporate latent variables and factors into the model and learn these variables and factors through optimization frameworks. External sources such as interaction networks or pathways are often used to construct the network or factor graph. For instance, PARADIGM \cite{Vaske2010} converted NCI pathway databases into a factor graph in which each gene is a factor incorporating several kinds of information. Since most of the variables and factors in this factor graph are unknown, PARADIGM relied on a set of empirical distributions to make the learning task possible with EM algorithm. 

Our approach presented in this report roughly falls into the third category -- non-probabilistic network approach. We do not assume a prior distribution for any variables to avoid unrealistic assumptions. Instead we adopted the techniques in spectral clustering to construct a k-nearest-neighbor affinity (similarity) graph with local Gaussian kernel (Eq.~\ref{eq:k_ij}). This will reduce noise and possible distortion caused by non-uniform measurements and preprocessing of omic datasets. Based on the main idea of similarity network fusion (SNF) \cite{Wang2014}, we developed a simpler and more general framework of affinity network fusion (ANF), or more technically speaking, transition matrix fusion, to combine multiple networks into a fused consensus network. The learned affinity network captures complementary information from multiple views and is much more robust than individual networks learned from each view.  

\section{General Framework to Cluster Complex Objects} \label{sec:general_framework}
\subsection{Representing Complex Objects With Multi-View}
A patient is a complex object, and can have multiple views with heterogeneous features. In this paper, we mainly deal with patient clustering. However, the proposed framework can be used for any complex object clustering.  

Suppose each patient has $n$-views. Thus we have $n$ patient-feature matrices:

$\mathcal{X}^{(v)} \in R^{N\times p_v}, v = 1,2,\cdots, n$

$N:$ number of patients

$p_v:$ feature dimension in view $v$

Here we assume each feature can be mapped to a vector of real numbers. Note that there are quite a lot of features that are not numeric, such as text annotations. However, many of them can be transformed to a new feature space in $R^p$ using feature embedding. In this paper, we do not focus on transforming a specific non-numeric feature space to $R^p$, but focus on clustering patients based on already transformed feature matrices in $R^{N\times p_v}$.

\subsection{Feature Selection and Transformation}
For patients, each type of omic data is different, representing a different view of the patients. The dimensionality of each view is usually very high. Feature selection and transformation are often necessary. 
For example, gene expression data can be  represented using a sample-feature matrix with tens of thousands of rows (i.e., genes) and dozens of columns (i.e., samples).
Most genes are not disease genes. If we use all gene features for clustering patients into subgroups, irrelevant gene features may lead to a ``bad'' clustering result. 

For cancer genomics, we usually have expression data from tumor-normal pairs, which can be used for selecting differentially expressed (DE) genes (or miRNAs). Using DE genes for clustering patients is often better than using all genes \cite{Ma2016}. 

In genomics, raw counts of molecular measurements are common. However, for clustering or other exploratory analysis, direct use of raw counts is usually discouraged. Proper feature normalization and transformation is often necessary before clustering. Log transformation, variance stabilizing transformation \cite{Durbin2002}, and regularized log transformation \cite{Love2014} are commonly used for transforming raw counts data for downstream analysis.

\subsection{Distance Metrics}
Most clustering techniques require to define a pair-wise patient distance (or similarity) matrix $\Delta = (\delta_{ij})_{N\times N}\in R_+^{N\times N}$.
$R_+$ represents the set of non-negative real numbers.

With proper feature engineering, we can calculate pair-wise distance on normalized features using Euclidean distance $\delta_{ij}=||x_i-x_j||$, or 
$\delta_{ij}=1-Cor(x_i, x_j)$ ($Cor(x_i, x_j)$ represents Pearson (or Spearman) correlation), etc. For categorical features without feature embedding, we can use chi-squared distance or other similar metrics.

\subsection{Clustering Objectives}
To find a clustering assignment function $\mathcal{A}=(a_1, a_2, \cdots, a_N)$ that maps each patient $i$ to a unique cluster $a_i$ (here we do not consider soft clustering) with certain constraints (e.g., the number of clusters), we need solve an optimization problem in general:

\begin{equation}
 \argmin_{\mathcal{A}} f(\mathcal{A}, \Delta)
\end{equation}

$f$ is a cost function that aims to make patients in the same cluster are more ``like'' each other than those belong to different clusters.

Once common choice is to minimize the ratio between the sum of intra-class distances and the sum of inter-class distances.

\begin{equation}
\argmin_{\mathcal{A}} \frac{\sum_{1\le i,j \le N}I(a_i = a_j)\delta_{ij}}{\sum_{1\le i,j \le N}I(a_i \ne a_j)\delta_{ij}}
\end{equation}

$I(\cdot)$ is the indicator function. Many clustering methods essentially solve this problem or its variants directly or indirectly.
If the distance between patients is not very accurate due to noise and feature heterogeneity, etc., we can build a similarity graph based on distance matrix $\Delta$, then perform graph cut to find densely connected clusters. A powerful approach is spectral clustering that work on patient similarity graph $S=(s_{ij})_{N\times N} \in R_+^{N\times N}$ \cite{Luxburg2007}. 
In this paper we used spectral clustering \cite{Luxburg2007} on learned fused affinity (similarity) matrix for patient clustering.

\section{kNN Affinity Network Fusion (ANF)} \label{sec:anf}
\subsection{Affinity Matrix for Each View}
With pair-wise distance matrix $\Delta$, we can define corresponding similarity graph $S$ in multiple ways. For example \cite{Luxburg2007}, 

\begin{itemize}
\item $\epsilon$ neighborhood: only the edges that has weight less than $\epsilon$ are kept in the similarity graph. The choice of $\epsilon$ is problem-dependent.

\item $k$-nearest-neighbor graph: only the edges of each node's k-nearest neighbors are kept. $k$ is the parameter to tune.

\item Fully connected graph: a kernel such as
$e^{-\frac{\delta_{ij}^2}{2\sigma^2}}$ is often used to transform distance to similarity. $\sigma$ is the radius of a local neighborhood instead of a global constant to capture local network structure.
\end{itemize}

In this paper, we adopted the definition of local $\sigma_{ij}$ from \cite{Wang2014}. 

\begin{equation}
\label{eq:mu_i}
\mu_i=\frac{\sum_{l\in N_k(i)}{\delta_{il}}}{k}
\end{equation}

\begin{equation}
\label{eq:sigma_ij}
\sigma_{ij}=\alpha(\mu_i+\mu_j) + \beta \delta_{ij}
\end{equation}

$N_k(i)$ represents the indexes of k-nearest neighbors of patient $i$. Thus $\mu_i$ in Eq.~\ref{eq:mu_i} represents local diameter of node $i$. $\sigma_{ij}$ in Eq.~\ref{eq:sigma_ij} incorporates both local diameters of patient $i$ and $j$ and their distance. The choice of $k$ is important and needs to be tuned. In \cite{Wang2014}, $\sigma_{ij}=\frac{\mu_i+\mu_j + \delta_{ij}}{3}$. Eq.~\ref{eq:sigma_ij} is more general with tuning parameters $\alpha$ and $\beta$, $\alpha, \beta \ge 0$. 

\begin{equation}
\label{eq:k_ij}
K_{ij}=\frac{1}{\sqrt{2\pi}\sigma_{ij}}e^{-\frac{\delta_{ij}^2}{2\sigma_{ij}^2}}
\end{equation}

Eq.~\ref{eq:k_ij} calculates local Gaussian kernel between patient $i$ and $j$, with $\sigma_{ij}$ defined as Eq.~\ref{eq:sigma_ij}, to incorporate local kNN network structure. Even though $K$ is fully connected (i.e., $\forall i, \forall j, K_{ij}>0$), only those node pairs that are within a small dense neighborhood will have a relatively large kernel (as similarity measure). We can regard $K$ as a similarity graph to perform spectral clustering.
 
With similarity matrix $K$, we can define a state transition matrix by Eq.~\ref{eq:s_ij}, with $S_{ij}$ representing the probability of (the state of) patient $i$ transition to (the state of) patient $j$. Each row of $S$ sums to 1. 

\begin{equation}
\label{eq:s_ij}
S_{ij} = \frac{K_{ij}}{\sum_{j=1}^{N}K_{ij}}, \quad 1 \le i,j \le N
\end{equation}

While $K$ is symmetric, $S$ is probably not. We use transition matrix instead of symmetric similarity matrix to make the our framework interpretable through random walk. 

\textbf{kNN Affinity Matrix for Each View}
With multi-view data, one can perform clustering on each view and synthesize results using approaches like consensus clustering \cite{Monti2003}. Or we can construct a more robust similarity network incorporating multi-view data and then perform spectral clustering.

For each view, we can calculate state transition matrix $S^{(v)}$ using Eq.~\ref{eq:s_ij}. Since $S^{(v)}$ is normalized from similarity matrix (Eq.~\ref{eq:k_ij}), we can easily recover a symmetric similarity graph from $S^{(v)}$. Thus we loosely refer $S^{(v)}$ as fully-connected similarity graph or affinity matrix in this paper.
 
Based on fully connected graph $S^{(v)}$, we can further define k-nearest-neighbor similarity graph or affinity matrix as $W^{(v)}$ (Eq.~\ref{eq:knn_transition}). 

\begin{equation}
\label{eq:knn_transition}
W_{ij}^{(v)}=
\begin{cases}
(1-\epsilon)\frac{S_{ij}^{(v)}}{\sum_{j\in N_k(i)}S_{ij}^{(v)}}, & \text{if}\ j \in N_k(i) \\
\epsilon \frac{S_{ij}^{(v)}}{\sum_{j\notin N_k(i)}S_{ij}^{(v)}}, & \text{otherwise}
\end{cases}
\end{equation}

$N_k(i)$ refers to the indexes of k nearest neighbors of patient $i$. $\epsilon$ refers to a small number. If we set $\epsilon = 0$, then for each row of $W^{(v)}$, only k elements are non-zero, and only the weights of k nearest neighbors are used for normalization. We also loosely call $W^{(v)}$ a (kNN) affinity matrix in this paper.

Since each row sums to 1, $W^{(v)}$ is also a transition matrix. In fact $W^{(v)}$ can be seen as a trunked version of $S^{(v)}$ by ``throwing away'' weak signals (i.e., small edge weights) in $S^{(v)}$. Thus $W^{(v)}$ should be more robust to small noise. If $W^{(v)}$ represents a connected and non-bipartite graph, it will reach a unique stationary distribution after a sufficient number of random walk \cite{Luxburg2007}.

When we cluster $N$ patients into several groups, we essentially try to find several different ``stable'' state space. Patients will be much more likely to stay in their own state space than to transition to another state space. Thus we can find a graph cut based on its transition matrix. For a network with multiple possible transition matrices from multi-view, we can use random walk on multigraph to aggregate all transition matrices to get a fused transition matrix for spectral clustering. This is the main idea of affinity network fusion (ANF) in this paper.

\subsection{Affinity Network Fusion with One-step Random Walk (ANF1)}

For each view, we have defined two transition matrices : $S^{(v)}$ (representing fully connected affinity network, Eq.~\ref{eq:s_ij}), and $W^{(v)}$ (kNN affinity network, a trunked version of $S^{(v)}$, Eq.~\ref{eq:knn_transition}).
We can build a multi-graph $\mathcal{G}$ to incorporate multi-views. In $\mathcal{G}$, each node represents a patient. There can be at most $n$ (or $2n$ if we include two edges from each view using Eq.~\ref{eq:s_ij} and Eq.~\ref{eq:knn_transition}) edges between patients. 

To calculate an aggregated edge weight for each patient pair, we can apply one-step random walk to fuse multi-view affinity networks in two steps.
First, we use Eq.~\ref{eq:anf1} to ``smooth'' each view. 
Then we use Eq.~\ref{eq:fused_mat} to get a fused weighted view.

\begin{equation}
\label{eq:anf1}
W^{(v)} = \beta_1 W^{(v)} + \beta_2 \overline{W^{(-v)}} + \beta_3 S^{(v)} + \beta_4 \overline{S^{(-v)}}
\end{equation}

\begin{center}
	$\sum_{v=1}^{4}\beta_v = 1, \beta_v \ge 0$
\end{center}

\begin{align}
\label{eq:complementary_view}
\overline{W^{(-v)}} &= \sum_{k\ne v}w_k\cdot W^{(k)}\\
\overline{S^{(-v)}} &= \sum_{k\ne v}w_k\cdot S^{(k)}
\end{align}

\begin{equation}
\label{eq:fused_mat}
W = \sum_{v=1}^{n}w_v\cdot W^{(v)}
\end{equation}

\begin{center}
	$\sum_{v=1}^{n}w_v = 1, w_v \ge 0$
\end{center}

In Eq.~\ref{eq:anf1}, the second term $\overline{W^{(-v)}}$ represents a weighted complementary view from $n-1$ other views (Eq.~\ref{eq:complementary_view}). 
Term 3 and term 4 in Eq.~\ref{eq:anf1} are included for comprehensiveness. In practice, since $W^{(v)}$ is usually more robust to noise than $S^{(v)}$, we often set $\beta_3 = \beta_4 = 0$. Eq.~\ref{eq:anf1} can be interpreted as network diffusion between view $v$ and other complementary views, resulting in a smoother version of $W^{(v)}$.

Since all $W^{(v)}$ and $S^{(v)}$ are transition matrices, the fused view Eq.~\ref{eq:fused_mat} essentially computes a weighted transition matrix $W$, which combines complementary information from multi-views and could be more informative for patient clustering. We can interpret the fused view $W$ (Eq.~\ref{eq:anf1} and \ref{eq:fused_mat}) as the result of one-step random walk on a multigraph, with $W$ being an aggregated transition matrix of a simple graph derived from multigraph. We call this process Affinity Network Fusion (ANF).
Even though it is very simple, it turned out to be as powerful as SNF \cite{Wang2014} (see Sec.\ref{sec:res}). 
 
To get an aggregated transition matrix, we can have multi-step random walk. In the following, we refer to ANF with one-step random walk as ANF1, and ANF with two-step random walk as ANF2, which is to be discussed in the next session. 

\subsection{Affinity Network Fusion with Two-step Random Walk (ANF2)}

In addition to one-step random walk, we can have multi-step random walk on multigraph. Our experiments on cancer genomic data showed a one-step or two-step random walk can usually work well enough. If the number of steps are too big, the fused transition matrix $W$ can eventually become rank 1, with each row being the same as the stationary distribution. Thus we only consider network fusion with two-step random walk in the following.

Similar to one-step random walk, we derive the fused transition matrix with two steps: first calculate a smoothed transition matrix for each view using Eq.~\ref{eq:anf2}, then aggregate all views using Eq.~\ref{eq:fused_mat}. 

\begin{align}
\label{eq:anf2}
\begin{split}
W^{(v)} = &\alpha_1 W^{(v)}\cdot \overline{W^{(-v)}} +
\alpha_2 \overline{W^{(-v)}} \cdot W^{(v)} +\\  
&\alpha_3 W^{(v)}\cdot \overline{S^{(-v)}} +
\alpha_4 \overline{S^{(-v)}} \cdot W^{(v)} + \\
&\alpha_5 S^{(v)}\cdot \overline{W^{(-v)}} +
\alpha_6 \overline{W^{(-v)}} \cdot S^{(v)} +\\ 
&\alpha_7 S^{(v)}\cdot \overline{S^{(-v)}} +
\alpha_8 \overline{S^{(-v)}} \cdot S^{(v)}
\end{split}
\end{align}
\begin{center}
	$\sum_{i=1}^{8}\alpha_i = 1, \alpha_i \ge 0$
\end{center}

The first term of Eq.~\ref{eq:anf2}, $\alpha_1 W^{(v)}\cdot \overline{W^{(-v)}}$ represents a two-step random walk (multiplying two transition matrices): the first step is a random walk on view $v$, the second step is a random walk on the aggregated complementary view ($\overline{W^{(-v)}}$, Eq.~\ref{eq:complementary_view}). Similarly, the second term represents random walk on the complementary view first followed by random walk on view $v$. Since we have two transition matrices $W^{(v)}$ and $S^{(v)}$, we can perform random walks on either $W^{(v)}$ or $S^{(v)}$. The other six terms in Eq.~\ref{eq:anf2} have similar meanings as the first two.

Our experiments on cancer genomic data show that the terms using $W^{(v)}$ usually works better than using $S^{(v)}$, suggesting $W^{(v)}$ is more reliable than $S^{(v)}$. In practice, the default choice is just using the first two terms:

\begin{equation}
\label{eq:anf2-simple}
W^{(v)} = \alpha W^{(v)}\cdot \overline{W^{(-v)}} +
(1-\alpha) \overline{W^{(-v)}} \cdot W^{(v)}
\end{equation}

\subsection{Comparison with SNF from \cite{Wang2014}}

ANF is based on SNF \cite{Wang2014}, but is much more simpler and as powerful as SNF. In SNF, the network fusion process is performed iteratively (Eq.~\ref{eq:snf}):

\begin{equation}
\label{eq:snf}
	S^{(v)} = W^{(v)} \times \frac{\sum_{k\ne v}S^{(v)}}{n-1} \times W^{(v)^T}
\end{equation}

Note the notations used in this paper are different from those in \cite{Wang2014}. In \cite{Wang2014}, $S^{(v)}$ represents a symmetric similarity matrix derived from Eq.~\ref{eq:k_ij}, while in our paper $S^{(v)}$ is a row-normalized asymmetric affinity (transition) matrix. Even though it is intuitive enough, SNF does not have a clear physical interpretation by multiplying similarity matrix ($S^{(v)}$) with transition matrices ($W^{(v)}$) in Eq.~\ref{eq:snf}. However, if we consider $S^{(v)}$ as a transition matrix, then Eq.~\ref{eq:snf} can be loosely seen as a three-step random walk. By contrast, in Eq.~\ref{eq:anf1} and Eq.~\ref{eq:anf2}, ANF directly operates on transition matrices, with a natural interpretation of random walk on multi-graph to generate an aggregated simple graph for spectral clustering. Our experiments showed that increasing the number of steps of random walk may not increase clustering performance dramatically. We have also tried iteratively updating $W^{(v)}$ as in \cite{Wang2014}. Results show that it is not necessary to include more iterations which often cannot outperform simple one-step or two-step random walk.  

In fact, if we adopt an iterative approach to update $W^{(v)}$, we have to ``manually'' adjust $W^{(v)}$ after each iteration. In order to ``force'' SNF to converge, \cite{Wang2014} used a ``mysterious'' operation to ``avoid numerical instability'' of $S^{(v)}$ by forcing the diagonal of $S^{(v)}$ to be 0.5 after each iteration (Eq.~\ref{eq:snf_mys}). As a result, the learned fused similarity matrix $S$ contains large values ($\approx 0.5$ based on their implementation) in the diagonal, while all other values are smaller by usually at least one order. Though the following spectral clustering does not rely on the diagonal elements, the physical meaning of the learned $S$ in SNF is not as clear as in ANF, where the learned $W$ is a weighted transition matrix.
	
\begin{equation}
\label{eq:snf_mys}
	S^{(v)}_{ij} = \begin{cases}
	\frac{1}{2}, \quad \text{if}\quad i=j\\
	\frac{1}{2}\frac{S^{(v)}_{ij}}{\sum_{j\ne i}S^{(v)}_{ij}} \quad \text{else}
	\end{cases} 
\end{equation}

Based on extensive experiments, we found this iterative process is not necessary. Without iterative process, the ``forced'' normalization (Eq.~\ref{eq:snf_mys}) can be eliminated, too.

Comparing ANF and SNF, ANF has at least several advantages:

First, it requires much less computation to achieve as good as or even better results than SNF (see Sec.~\ref{sec:res}). SNF typically requires about dozens of iterations to converge, while ANF only needs no iteration. ANF1 (Eq.~\ref{eq:anf1}) does not involve matrix multiplication, ANF2 (Eq.~\ref{eq:anf2-simple}) involves two matrix multiplications, while SNF (Eq.~\ref{eq:snf}) needs perform two matrix multiplications for each iteration. 

Second, ANF is more general framework that can incorporate weights of views, while SNF only use uniform weights (Eq.~\ref{eq:snf}). This is important because properly chosen weights can make fusion process more effective.

Third, ANF has a natural interpretation of random walk on multi-graph to generate a fused simple graph. The learned fused affinity matrix $W$ has a natural meaning of weighted transition matrix incorporating multi-view data, while the operation in SNF does not have a direct physical meaning, though it can also be loosely seen as a three-step random walk. 

To sum up, ANF significantly reduces unnecessary operations in SNF, and provides a more general and interpretable framework for integrating multi-view data using patient network fusion. 

\subsection{Spectral Clustering on Fused Affinity Matrix} \label{sec:spec}
With learned fused affinity matrix $W$, we can perform spectral clustering by solving an optimization problem Eq.~\ref{eq:optim_spec} \cite{Stella2003}.
\begin{align}
\label{eq:optim_spec}
\begin{split}
\argmax_Y \quad Trace((Y^TDY)^{-\frac{1}{2}}Y^TWY(Y^TDY)^{-\frac{1}{2}})\\
Y\mathbf{1}_K=\mathbf{1}_N, Y\in \{0,1\}^{N\times K}
\end{split}
\end{align}

$D$ is a diagonal matrix with diagonal elements being the sum of each row in $W$. Since $W$ generated by ANF is a transition matrix, $D$ becomes an identity matrix. Eq.~\ref{eq:optim_spec} finds clustering assignment matrix $Y$ that maximizes K-way normalized associations \cite{Stella2003}, which can be solved approximately by solving Eq.~\ref{eq:optim2_spec} \cite{Stella2003}.
\begin{align}
\label{eq:optim2_spec}
\begin{split}
\argmax Trace(Y^TZR) \\
D^{-1}WZ=Z\Lambda \\
R^TR=I_K, Y\mathbf{1}_K=\mathbf{1}_N, Y\in \{0,1\}^{N\times K}
\end{split}
\end{align}

The columns of $Z$ in Eq.~\ref{eq:optim2_spec} are eigenvectors of $D^{-1}W$ (or $W$ since $D$ is identical matrix). $R$ is orthogonal matrix.
We solve Eq.~\ref{eq:optim2_spec} by iteratively update $R$ and $Y$ \cite{Stella2003}. 
First fix $R$, update $Y$
\begin{equation}
Y_{ij}=I(j=argmax(ZR)_{ij})
\end{equation}

$I(\cdot)$ is indicator function.
Then fix $Y$, update $R$
\begin{align}
Y^TZ=U\Lambda V^T\\
R = VU^T
\end{align}
$Y^TZ=U\Lambda V^T$ is the singular value decomposition of $Y^TZ$.

The overall ANF framework to cluster cancer patients is summarized in Alg.~\ref{alg:ANF_framework}.

\begin{algorithm}
	\label{alg:ANF_framework}
	\SetKwInOut{Input}{Input}
	\SetKwInOut{Output}{Output}
	
	\Input{\textbullet{Patient-feature matrices ($n$ views): $\mathcal{X}^{(v)},v=1,2,\cdots,n$} \\ 
		\textbullet{Number of clusters: $K$}\\
		\textbullet{Weight of each view (optional): $w$} \\
		\textbullet{Other optional parameters, such as weight of ANF components $\alpha$}}
	\Output{
		\textbullet{Fused patient affinity matrix $W$}\\
		\textbullet{Patient cluster assignment $Y$}}
	\Begin{\textbf{Feature selection and transformation}\\
		$\mathcal{X}^{(v)} \rightarrow X^{(v)}\in R^{N\times p_v},v=1,2,\cdots,n$\\ 
		\textbf{Calculate pair-wise distance matrix for each view: $\Delta^{(v)}\in R_+^{N\times N},v=1,2,\cdots,n$}\\
		\textbf{Calculate kNN affinity matrix for each view: $W^{(v)},v=1,2,\cdots,n$} (Eq.~\ref{eq:anf1} or Eq.~\ref{eq:anf2})\\
		\textbf{Calculate fused affinity matrix $W$} (Eq.~\ref{eq:fused_mat})\\	
		\textbf{Spectral clustering on fused affinity matrix $W$: $(W,K)\rightarrow Y$} (Sec.~\ref{sec:spec})\\
		\textbf{Return $W, Y$}
	}	
	\caption{Affinity Network Fusion for Patient Clustering}
\end{algorithm}

\section{Experimental Results} \label{sec:res}

\subsection{Dataset and Evaluation Metrics}
Harmonized cancer datasets were downloaded from Genomic Data Commons Data Portal (\url{https://portal.gdc.cancer.gov/}).
We selected four primary sites with more than one disease type: adrenal gland, lung, kidney, and uterus. For example, cancers from adrenal gland has two disease types: Pheochromocytoma and Paraganglioma (project name: TCGA-PCPG) and Adrenocortical Carcinoma (project name: TCGA-ACC). In this paper, for ease of description, we refer to ``cancer types'' as cancers from these four primary sites. We want to cluster tumor samples of the same ``cancer types'' into known disease types. 
The number of samples used for analysis in each cancer type is summarized in Table~\ref{tbl:sample_info_4types} (a few ``outlier'' samples detected by exploratory data analysis had already been removed). 
All these patient samples has gene expression, miRNA expression and DNA methylation (from HumanMethylation450 array) data available for both tumor and normal samples.   

\begin{table}[t]
	\begin{center}
	\caption{Sample information of four cancer types}
	\label{tbl:sample_info_4types}
	\begin{tabular}{c|cc|c}
		\firsthline
		Cancer type &  \multicolumn{2}{c|}{Disease type} & Total \\
		\hline
		\multirow{2}{*}{adrenal gland} & TCGA-ACC & 76 & \multirow{2}{*}{253}\\
		& TCGA-PCPG & 177 &\\
		\hline
		\multirow{2}{*}{lung} & TCGA-LUAD & 447 & \multirow{2}{*}{811}\\
		& TCGA-LUSC & 364 &\\
		\hline
		\multirow{3}{*}{kidney} & TCGA-KICH & 65 & \multirow{3}{*}{654}\\
		& TCGA-KIRC & 316 & \\
		& TCGA-KIRP & 273 & \\
		\hline
		\multirow{2}{*}{Uterus} & TCGA-UCEC & 421 & \multirow{2}{*}{475}\\
		\cline{2-3}
		& TCGA-UCS  & 54 & \\
		\lasthline
	\end{tabular}
\end{center}
\end{table}

While our ultimate goal is to detect cancer subtypes (the true subtypes are not known yet), it is a good strategy to evaluate disease subtype discovery methods using a dataset with groundtruth. The dataset we selected and processed serves for this purpose well.  

For tumors from each primary site, we already know the disease types. Since we have ground truth disease types, we can evaluate clustering results using external metrics such as normalized mutual information (NMI). In addition, we have cancer patient survival data. We can perform survival analysis to test if the patient clusters show statistically different survival distributions.

The three metrics we used to evaluate clustering results are: 
(1) Normalized mutual information: 
$NMI(\Omega, \mathcal{C})=\frac{I(\Omega, \mathcal{C})}{(H(\Omega)+H(\mathcal{C}))/2}$;
(2) Adjusted Rand Index (ARI) \cite{Hubert1985},  
and (3) p-value of log rank test of survival distributions of different patient clusters \cite{HARRINGTON1982}.

We have chosen seven combinations of data types (legend of Fig.~\ref{fig:power_ANF}) and six feature types of gene expression and miRNA expression (legend of Fig.~\ref{fig:power_feature}). For DNA methylation, we directly used beta values, so it only has one feature type. Thus in total there are 37 unique combinations of data types and feature types. We run both ANF1 and ANF2 on all 37 combinations, and SNF on 24 combinations (ANF is implemented to work on a single data type as well, while the implementation of SNF requires input to include at least two data types). Due to page limit, detailed results including code can be accessed at \url{https://github.com/BeautyOfWeb/ANF}. In the following, we only show some results to demonstrate the power of ANF and feature engineering, and compare ANF with SNF.

\subsection{The Power of Affinity Network Fusion (ANF)}
\begin{figure}[!t]
	\centering
	\includegraphics[width=2.5in]{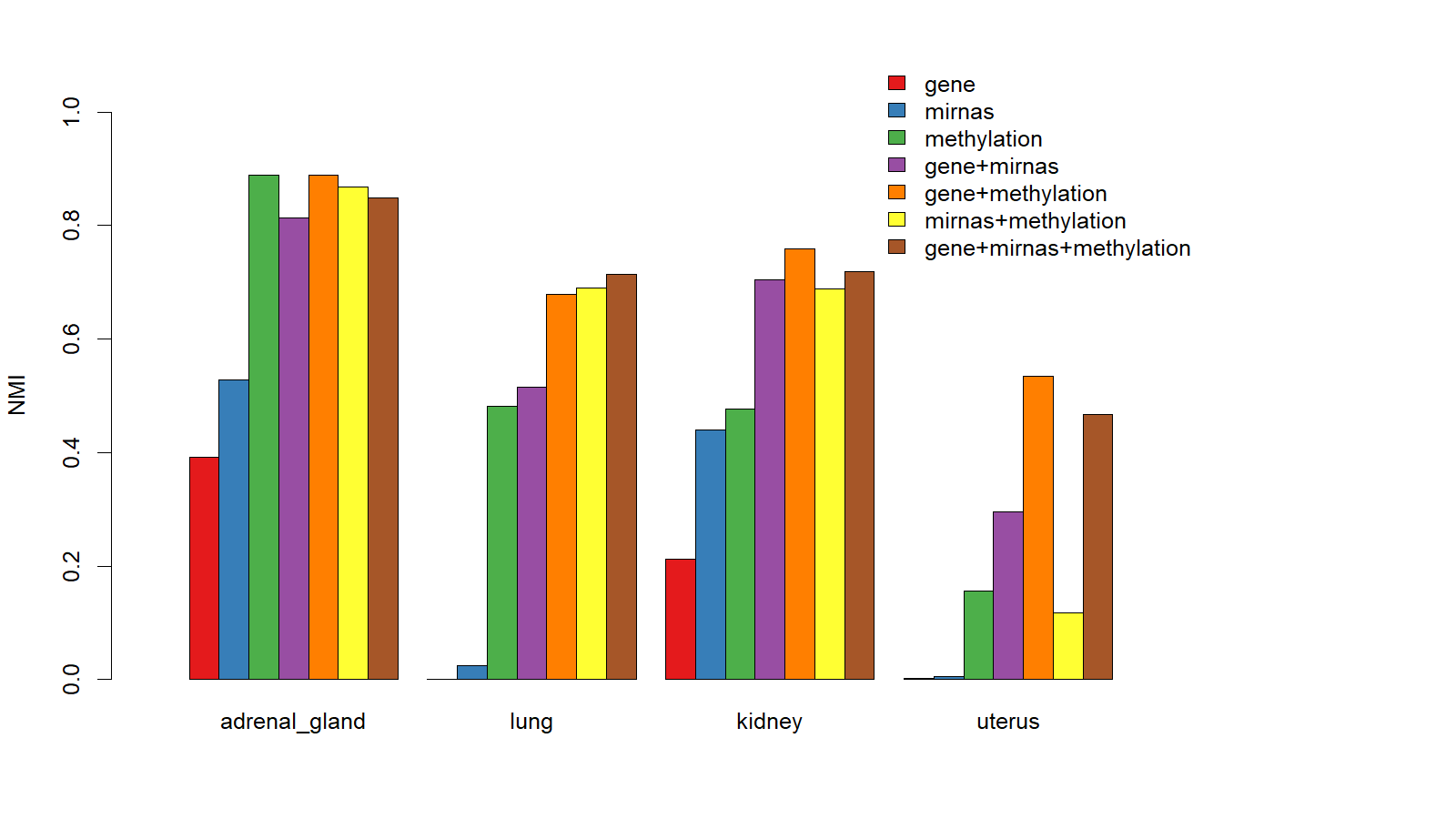}
	\caption{Power of ANF combining multiple data types}
	\label{fig:power_ANF}
\end{figure}

To demonstrate the power of ANF, we compared the clustering results using single data types with those using ANF to integrate multiple data types.
In Fig.~\ref{fig:power_ANF}, we compared seven different combinations of data types: 
\begin{itemize}
	\item ``gene'': gene expression
	\item ``mirnas'': miRNA expression
	\item ``methylation'': DNA methylation (beta values from Illumina Human Methylation 450 platform)
	\item ``gene+mirnas'': combine ``gene'' and ``mirnas'' using ANF
	\item ``gene+methylation'': combine ``fpkm'' and ``methylation'' using ANF
	\item ``mirnas+methylation'': combine ``mirnas'' and ``methylation'' using ANF
	\item ``gene+mirnas+methylation'': combine ``gene'', ``mirnas'', and ``methylation'' using ANF
\end{itemize}

Fig.~\ref{fig:power_ANF} shows NMI values (between 0 and 1. The larger NMI value, the better clustering result) of patient clusters (here we set the number of clusters to be the number of disease types) using ANF2 framework on the aforementioned seven combinations of data types (for gene expression, we used normalized FPKM values; for miRNA expression, we used normalized counts in Fig.~\ref{fig:power_ANF}). Fig.~\ref{fig:power_ANF} shows clustering using DNA methylation beta values performs better than using FPKM and normalized miRNA expression values for all four cancer types, suggesting that DNA methylation data may contain highly relevant information about disease types (and potentially disease subtypes). 

In general a combination of at least two data types usually yields better clustering results. There are two exceptions in Fig.~\ref{fig:power_ANF}. For adrenal gland cancer, clustering using DNA methylation beta values alone yields the best clustering results (``gene+methylation'' can also generate the same result), and can outperform the results from using a combination of data types. Again this suggests DNA methylation beta values contain most relevant information about disease (sub)types. Another exception is that the clustering result using ``mirnas+methylation'' combination for uterus cancer is lower than using ``methylation'' data alone. This is probably due to that the quality of fused affinity network from miRNA and methylation data is not as good as that from methylation data alone. 
For uterus cancer, clustering using gene or miRNA expression data alone did a ``terrible'' job (NMI $\approx 0$). However, by integrating the two data types, the result improves significantly (NMI=0.30), which demonstrates the power of ANF. 

We also find that it is usually not the case that integrating three data types would generate better results than that from integrating two data types. However, integrating more data types tends to make the results more robust as ``gene+mirna+methylation'' consistently performs relatively well across four cancer types, while two data type combinations may fail in some cases. For instance, for uterus cancer, integrating miRNA and methylation data does not perform well, while integrating all three types of data performs much better even though it does not achieve the best result. Note in Fig.~\ref{fig:power_ANF}, we used ANF2 and did not tune the weight of views. Better results may be achieved if we tune the weights.

\begin{figure}[!t]
	\centering
	\includegraphics[width=2.5in]{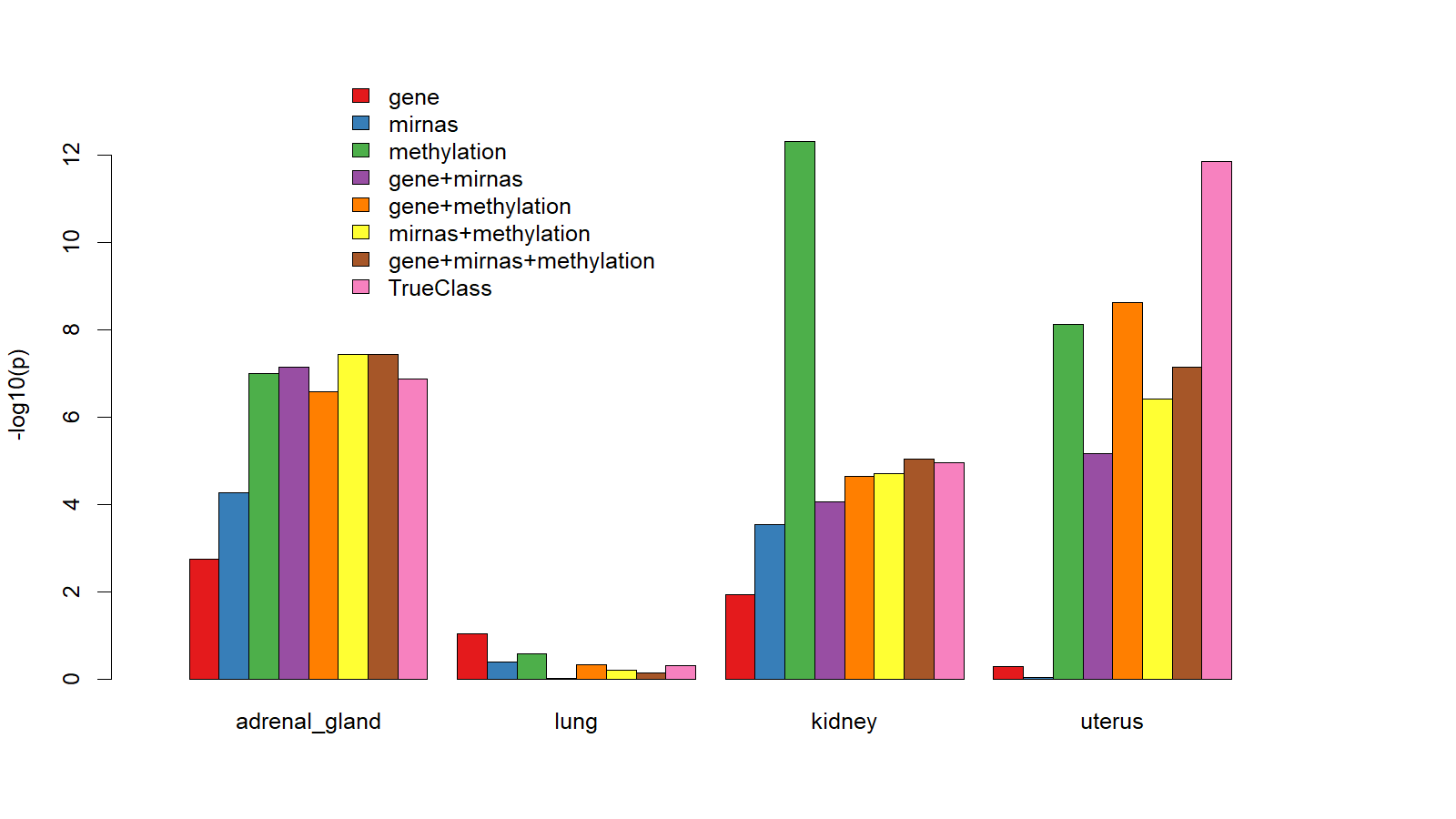}
	\caption{-log(p-value) of log rank test of patient survival distributions}
	\label{fig:power_ANF_pval}
\end{figure}

Very similar results are obtained for using Adjust Rand Index (ARI) as clustering metric (not shown here). Fig.~\ref{fig:power_ANF_pval} shows $-log_{10}(p\textnormal{-}value)$ of log rank test of patient survival distributions among detected patient clusters. Since we already know the true disease labels, we added a pink bar labeled ``TrueClass'' in Fig.~\ref{fig:power_ANF_pval} (the other seven bars are one-to-one matched with Fig.~\ref{fig:power_ANF} for comparison). Label ``TrueClass'' does not correspond to a data type combination, but refers to the groundtruth cluster assignment (the bar corresponding to ``TrueClass'' shows the negative log p-value of log-rank test of survival distributions of true disease types). 

Fig.~\ref{fig:power_ANF_pval} shows it is not always the case that the p-value calculated from using true class labels is the smallest (negative log p-value the largest). In fact, for lung cancer, the survival distributions of two known disease types do not shown statistical difference at all, even though we can separate the two disease types relatively well using clustering. For kidney cancer, the smallest p-value ($p=4.7\times 10^{-13}$) was achieved using methylation data alone for clustering, but the corresponding clustering accuracy was relatively low (NMI=0.48, Fig.~\ref{fig:power_ANF}). Using the true class labels of kidney cancer, we can only achieve p-value = $1.1\times 10^{-5}$.
This suggests that log-rank test of survival distributions should not be used as the only metric to evaluate patient clustering results. This is also one major reason we carefully select such a dataset with groudtruth disease type information for evaluation purpose. With groudtruth disease type information, external evaluation metrics such as NMI and Adjusted Rand Index (ARI) can be used. The largest possible value of NMI and ARI is 1 if all patients are clustered correctively.

\subsection{The Power of Feature Selection and Transformation}
\begin{figure}[!t]
	\centering
	\includegraphics[width=2.5in]{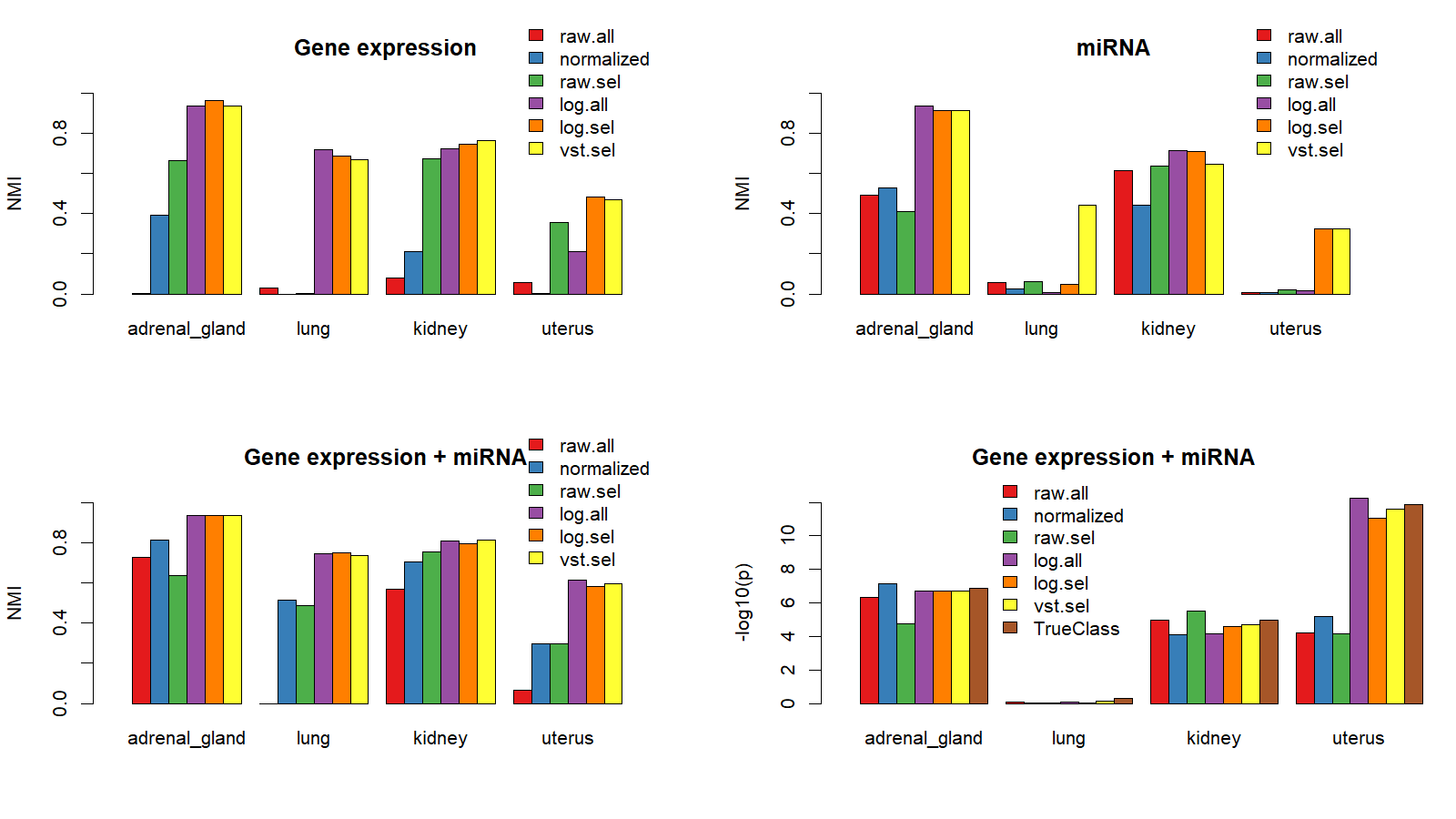}
	\caption{Power of feature engineering}
	\label{fig:power_feature}
\end{figure}

Fig.~\ref{fig:power_feature} shows the power of feature selection and transformation of raw counts data (gene and miRNA expression). The topleft, topright, and bottomleft panels show NMI of clustering results using gene expression, miRNA expression, and a combination of both using ANF. We choose gene and miRNA expression data because raw counts data are available for both data types, and we can apply commonly used feature selection and transformation techniques to raw counts. We used differential expression analysis to select gene or miRNAs features, and used two commonly raw counts transformation techniques: 
1) log transformation: $log_2(n+n_0)$. 
(in our experiments, we set $n_0=1$, so that zero counts will be mapped to 0), and 2)
variance stabilization transformation \cite{Durbin2002}. 

In Fig.~\ref{fig:power_feature} we compared clustering results using six different features. 
\begin{itemize}
	\item ``raw.all'': Raw counts of all genes or miRNAs
	\item ``normalized'': FPKM values of all genes or normalized counts for all miRNAs
	\item ``raw.sel'': Raw counts of selected (differentially expressed) genes or miRNAs (Differential expression analysis was performed using DESeq2 \cite{Love2014})
	\item ``log.all'': Log transformation of raw counts of all genes or miRNAs
	\item ``log.sel'': Log transformation of raw counts of selected (differentially expressed) genes or miRNAs
	\item ``vst.sel'': Variance stabilizing transformation of raw counts of selected genes or miRNAs
\end{itemize}

The topleft panel of Fig.~\ref{fig:power_feature} shows that for all four cancer types, ``log.sel'' and ``vst.sel'' of gene expression will work relatively well. The topright panel shows that ``vst.sel'' of miRNA expression consistently works relatively well across all four cancer types. When combining both gene expression and miRNA expression data using ANF (bottomleft panel), ``log.all'', ``log.sel'', and ``vst.sel'' work consistently well. For each type of feature, combining two data types using ANF improves clustering accuracy. This also demonstrates the power of ANF, consistent with the analysis of Fig.~\ref{fig:power_ANF}. 

``TrueClass'' was only shown in bottomright panel to compare the negative log p-value of log-rank test of survival distributions when true disease labels are used. Some clustering results have a smaller p-value than the p-value calculated using true patient groups even when the clustering results are not 100\% correct (Fig.~\ref{fig:power_feature} bottomleft and bottomright panels). This is consistent with the analysis of Fig.~\ref{fig:power_ANF_pval}. 

\subsection{Performance Comparisons with SNF from \cite{Wang2014}}

\begin{figure}[!t]
	\centering
	\includegraphics[width=2.5in]{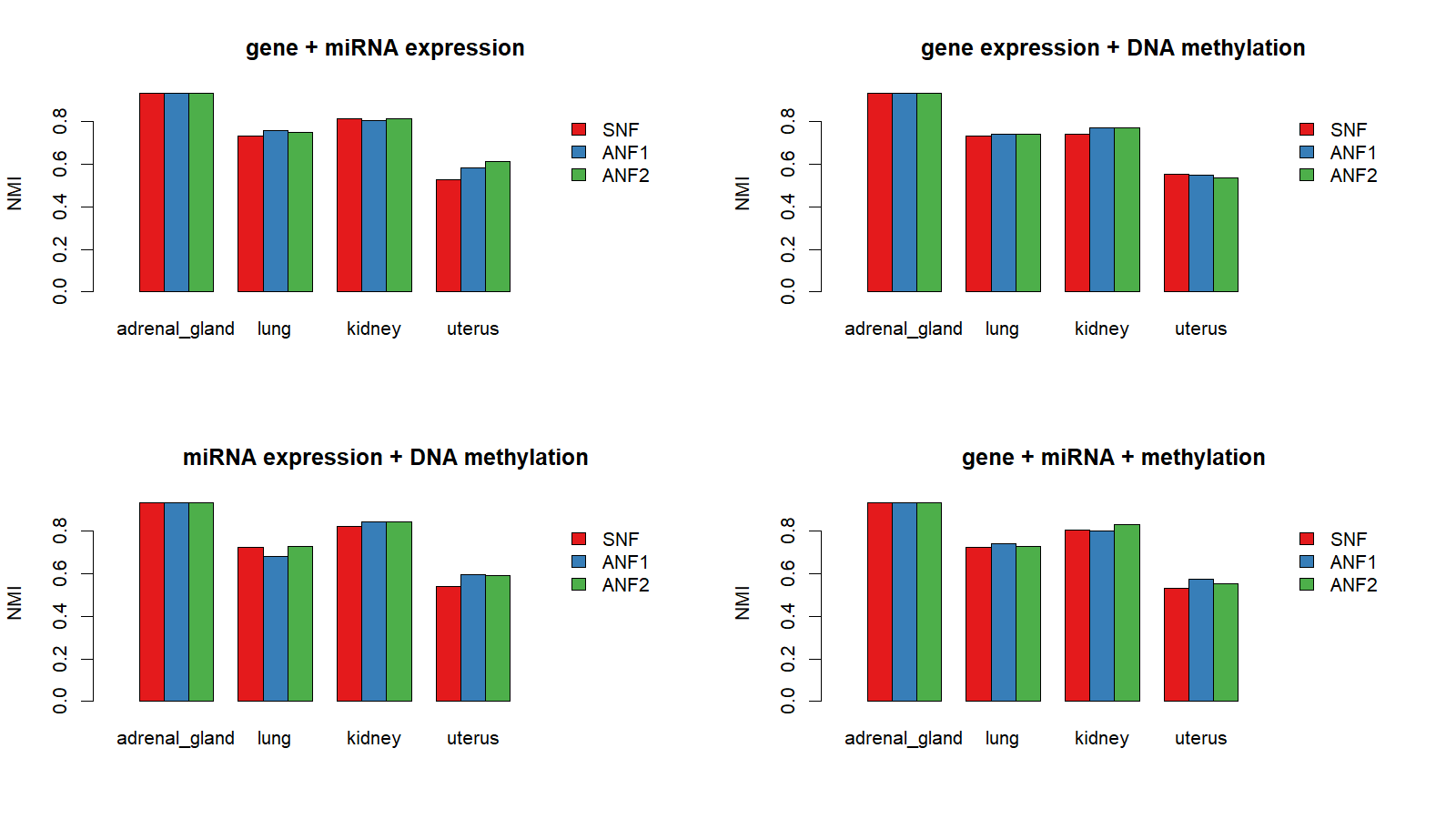}
	\caption{Comparing the performance of SNF and two frameworks of ANF}
	\label{fig:ANF_SNF_3types}
\end{figure}

Fig.~\ref{fig:ANF_SNF_3types} compares the performances of SNF, ANF1 and ANF2 for clustering four cancer types into their known disease types. Since SNF has shown superior performance over KMeans, iCluster, and feature concatenation approaches \cite{Wang2014}, we do not repeat comparisons with KMeans, iCluster, etc. in this paper.

We have used three types of data: gene expression, miRNA expression and DNA methylation. The topleft, topright, bottomleft, and bottomright panels correspond to results from four different combinations of data types.

Except adrenal gland, for which all three methods can achieve the same clustering accuracy (NMI=0.93, only two out of 253 samples were misclassified), at least one of ANF1 and ANF2 can achieve slightly better results than SNF.

Though the improvement is not significant, ANF has several advantages. ANF has a much more transparent interpretation and uses much less computation compared with SNF. SNF needs to converge after typically 20 iterations, while ANF only needs less than half computation required for computing one single iteration of SNF. Moreover, ANF provides a more general framework and can incorporate the weights of multiple views, while SNF only used uniform weights. Taken these facts into consideration, ANF is better than SNF and thus can replace SNF for clustering complex patients with multi-view data. 

There is no significant difference using ANF1 and ANF2, which correspond to one-step and two-step random walk on a multigraph to generate a fused simple graph for spectral clustering, respectively.

Both ANF1 and ANF2 has weight parameters for each view. We found uniform weights can usually do a good job. If we set the weight of each view to be the NMI value of the clustering result using that view alone, we can usually get a slightly better result. However, the optimal weights may not be very intuitive as simple NMI values. In our experiments (Fig.~\ref{fig:power_ANF}), we randomly set weights as $(\frac{1}{6},\frac{1}{3}, \frac{1}{2})$ for gene expression, miRNA expression and DNA methylation, and can achieve slightly better results using NMI values as weights in some cases. We can improve results by tuning weights of each view, however, this usually requires using true class label information. In an unsupervised task (our framework ANF is designed for unsupervised learning, we just used true class labels to evaluate its performance), we have to use internal evaluation metric such as silhouette or resort to some related information (such as cancer patient survival data) to evaluate clustering results.

In addition, since ANF framework applied spectral clustering to a fused affinity matrix, we can use eigengap heuristic to determine the number of clusters and indirectly assess cluster quality. 

\subsection{Determine the Number of Clusters Using Eigengap Analysis} \label{sec:eigengap}

\begin{figure}[!t]
	\centering
	\includegraphics[width=2.5in]{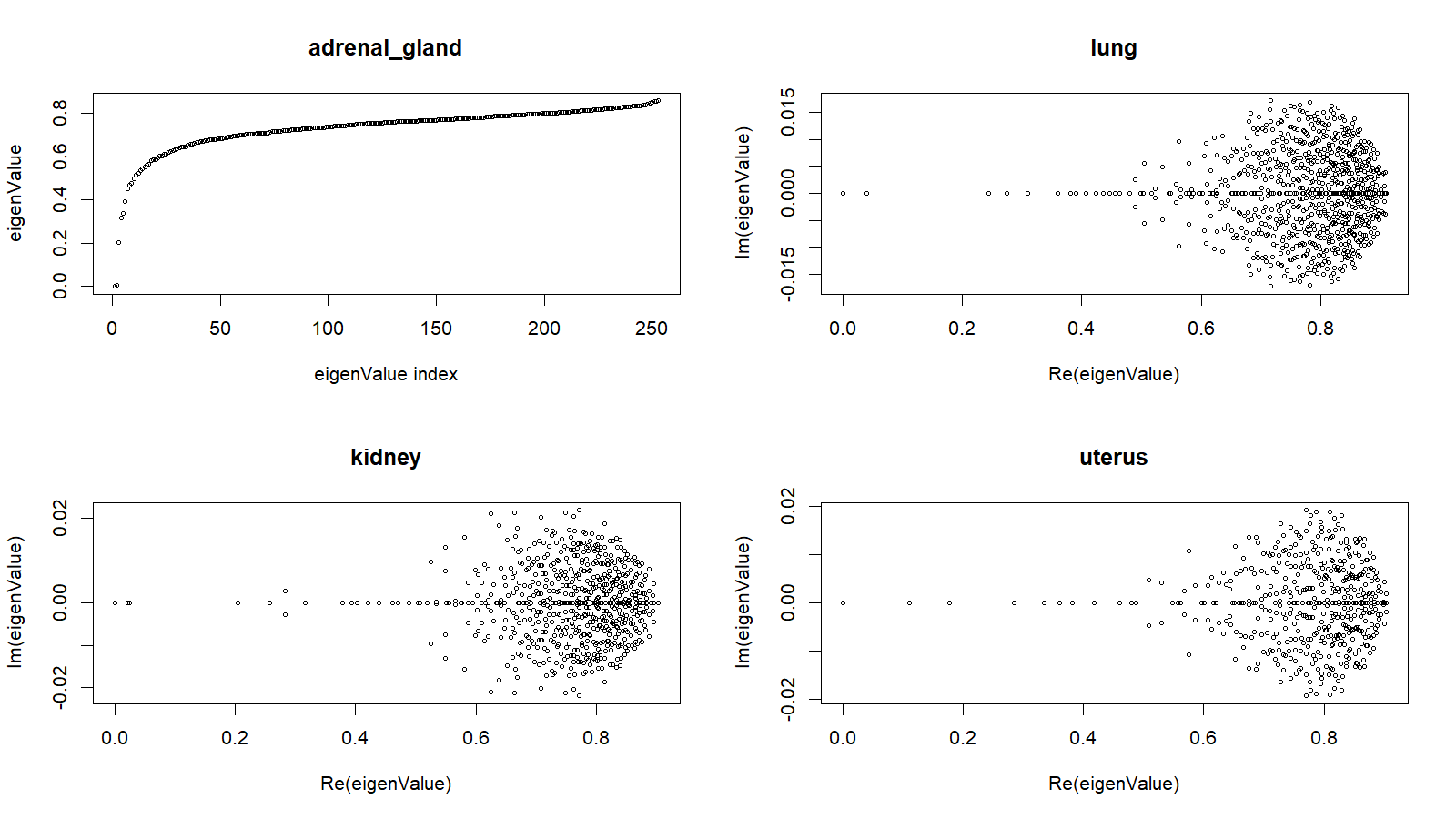}
	\caption{Eigenvalues of affinity matrix of four cancer types}
	\label{fig:eigenValues_4cancers}
\end{figure}

In this section, we carefully examine the learned fused affinity matrices for the four cancer types. We have run ANF on 37 combinations of features types and data types, We choose the affinity matrix $W$ with the highest NMI value for each cancer type for the following analysis. 

The fused affinity matrix $W$ generated by ANF is a transition matrix, and is asymmetric in most cases if the node degree distribution is non-uniform. We apply spectral clustering on this affinity network with normalized graph Laplacian $L$ being:

\begin{equation}
\label{eq:graph_laplacian}
	 L=I-D^{-1}W
\end{equation}
 
$D$ is a diagonal matrix with diagonal elements being the row sums of $W$. Though $L$ is asymmetric, experiments show that the results can be as good as or slightly better than a symmetric version of $L$ (one way to get a symmetric $L$ is first force $W$ to be symmetric $W\leftarrow(W+W^T)/2$, then calculate $L$). Let's define asymmetric ratio of $W$ as:

\begin{center}
$AsymRatio(W)=\frac{||W-W^T||^2}{||W||^2}$
\end{center}

$W^T$ is the transpose of $W$, and $||\cdot||$ represents Frobenius norm. The asymmetric ratios of the fused affinity (transition) matrices $W$ learned by ANF are 0, 0.10, 0.11, 0.11 for adrenal gland, lung, kidney and uterus, respectively.
The eigenvalues of the corresponding normalized graph Laplacian $L$ are shown in Fig.~\ref{fig:eigenValues_4cancers}.

Surprisingly, for adrenal gland, the learned affinity matrix $W$ is symmetric (with all eigenvalues being real numbers shown in topleft panel of Fig.~\ref{fig:eigenValues_4cancers}). With this $W$, all 253 samples except one are clustered correctly according true diseae types (NMI=0.96). For other cancer types, $W$ is not asymmetric and there are quite a few complex eigenvalues. However, the imaginary part of eigenvalues are less than 0.02, and the real part of eigenvalues are at least more than 30 times larger than the imaginary part. Let's use perturbation theory to briefly explain this observation.

We can treat an asymmetric $L$ as a symmetric matrix plus a perturbation matrix (representing small disturbances):

\begin{center}
 $L = L_{sym}+H$
\end{center}
 
Based on perturbation theory, as long as $H$ is relatively ``small'' than $L_{sym}$, the eigenvalues and corresponding eigenvectors of $L$ should be near those of $L_{sym}$. The eigenvalues of $L_{sym}$ are all real numbers with the smallest eigenvalue being 0 ($0\le \lambda_1 \le \lambda_2 \le \cdots \lambda_n$), and thus the eigenvalues of $L$ should be approximately the same as those of $L_{sym}$.

One interesting property of our defined $W$ and $L$ is the summation of eigenvalues of $L$ equals to the summation of the diagonal of $L$, which equals to:
\begin{center}
$\sum_{i=1}^{N}\lambda_i = N-\sum_{i=1}^{N}W_{ii}$
\end{center}

We observed all eigenvalues (absolute value for complex value) are in [0,1]. We expect this to be generally true for any transition matrix $W$, and thus for $L$ (without rigorous proof). 

Importantly, we found eigengap heuristic is very useful for deciding the number of clusters. For adrenal gland, the first two smallest eigenvalues are very near 0, while the third one is about 0.2. The eigengap between the third and second smallest values is relatively large. This suggests there should be two ``natural'' clusters (corresponding to the two nearly 0 eigenvalues). Furthermore, the eigengap between the fourth and third values is relatively high, too. This suggests we can use the learned affinity matrix $W$ for disease subtype discovery for adrenal gland. In fact, when we set the number of cluster to be 3, our framework will separate 176 ``TCGA-PCPG'' samples into two groups consisting 155 samples and 21 samples respectively. If we set the number of clusters to 4, the 155 samples will be further split into two small groups as shown in Table.~\ref{tbl:confusion_adrenal}.

\begin{table}[t]
	\begin{center}
	\caption{Confusion matrix of clustering adrenal gland}
	\label{tbl:confusion_adrenal}
\begin{tabular}{c|c|ccccc}
	\firsthline
	\#Clusters & TrueClass & \multicolumn{5}{c}{Clusters} \\
	\hline
	\multirow{3}{*}{2} & & C1 & C2 \\
	\cline{3-4}
 & TCGA-ACC & 0 & 76 \\ 
 \cline{3-4}
 & TCGA-PCPG & 176 & 1 \\
 \hline
 \multirow{3}{*}{3} & & C1 & C2 & C3\\
 \cline{3-5}
 & TCGA-ACC & 0 & 0 & 76\\ 
 \cline{3-5}
 & TCGA-PCPG & 155 & 21 & 1\\
 \hline
\multirow{3}{*}{4} & & C1 & C2 & C3 & C4\\
\cline{3-6}
& TCGA-ACC & 0 & 0 & 76 & 0\\ 
\cline{3-6}
& TCGA-PCPG & 83 & 21 & 1 & 72\\
 \hline
\multirow{3}{*}{5} & & C1 & C2 & C3 & C4 & C5\\
\cline{3-7}
& TCGA-ACC & 0 & 0 & 30 & 46 & 0\\ 
\cline{3-7}
& TCGA-PCPG & 83 & 21 & 0 & 1 & 72\\
	\lasthline
\end{tabular}
\end{center}
\end{table}

Similarly, for lung cancer, the two smallest eigenvalues are near zero, and the eigengap between the third and second values is relatively high. This information alone suggests there should be two ``natural'' clusters, which coincide with the two different disease types. 

For kidney cancer, three eigenvalues are near 0 while the eigengap between the fourth and the third one is relatively big. From this information, we can infer that there should be three ``natural'' clusters, which coincide with the fact that there are indeed three different disease types in kidney cancer. Our clustering results can achieve high accuracies for adrenal gland, lung, and kidney cancers (Table~\ref{tbl:cluster_accuracy_4types}).

For uterus cancer, only one eigenvalue is 0. The eigengap between the second and first value is already relatively large. Thus there might not be ``natural'' clusters (``natural'' clusters emerge from some nearly block diagonal affinity matrix $W$). Consequently, our clustering result only achieves a moderate accuracy with NMI = 0.61 and adjusted rand index = 0.78. However, the p-value of log-rank test of survival distributions of two clusters is $5.4 \times 10^{-13}$, while the p-value calculated from the true disease types is $1.38\times 10^{-12}$. This suggests the identified clusters may still be useful to define clinically different patient groups.

\begin{table}[t]
	\begin{center}
	\caption{Clustering accuracy of four cancer types}
	\label{tbl:cluster_accuracy_4types}
	\begin{tabular}{c|c|c|c|c}
		\firsthline
	 & Adrenal gland & Lung & Kidney & Uterus \\
	 \hline 
	 NMI & 0.96 & 0.75 & 0.84 & 0.61\\
	 \hline
	 ARI & 0.98 & 0.83 & 0.91 & 0.78 \\
	\hline
	$-log_{10}(p)$ & 6.8 & 0.04 & 4.76 & 12.3\\
	\lasthline
	\end{tabular}
	\end{center}
\end{table}

Without true class label information, eigengap analysis can be used to predict the number of clusters and assess the ``cluster quality'' of affinity matrix for spectral clustering. In fact, the above eigengap analysis of the learned affinity matrices successfully reveal the potential number of ``natural'' clusters and is consistent with spectral clustering theory. This suggests the four learned fused affinity matrices may have successfully captured patient group structure, and can be used for unknown cancer subtype detection (For example, we can further cluster adrenal gland into more than two clusters as shown in Table.~\ref{tbl:confusion_adrenal}).

\section{Conclusion}

Defining cancer subtypes and identifying subtype-specific molecular signatures associated with clinical variables is one major goal for cancer genomics. 
In this paper, we presented affinity network fusion (ANF) framework, an upgrade of SNF \cite{Wang2014}, for clustering cancer patients by integrating multi-omic data. ANF has a clear interpretation, is more general than SNF, and can achieve as good as or even better results than SNF with much less computation. 
We performed extensive experiments on a selected cohort of 2193 cancer patients from four primary sites and nine disease types, and achieved high clustering accuracy. With this carefully selected ``gold'' dataset, we demonstrated the power of ANF and the power of feature selection and transformation in cancer patient clustering. 

Eigengap analysis on learned fused affinity matrices is highly consistent with true class label information, which strongly suggests that the learned affinity matrices may capture the internal structure of patient groups. We can use these matrices for subsequent cancer subtype discovery. Once disease subgroups are defined, future work may focus on a relatively homogeneous group of patients to identify subtype-specific comprehensive molecular signatures. 

While we only reported experimental results on four cancer types with known disease types, ANF can be used for discovering subtypes of other cancers, and more generally, for complex object clustering with multi-view feature matrices.

\bibliographystyle{IEEEtran}
\bibliography{IEEEabrv,F:/References/Bib/network-integration}

\end{document}